\documentstyle[12pt]{article}
\renewenvironment{thebibliography}[1]                   
        {
         \begin{list}{\arabic{enumi}.}                  
        {\usecounter{enumi}\setlength{\parsep}{0pt}
         \setlength{\leftmargin 17pt}{\rightmargin 0pt} 
         \setlength{\itemsep}{0pt} \settowidth          
        {\labelwidth}{#1.}\sloppy}}{\end{list}} 

\newfont{\spec}{msbm10 scaled 1200}

\newcommand{\be}{\begin{eqnarray}}
\newcommand{\ee}{\end{eqnarray}}

\newcommand{\cN}{{\cal N}}
\newcommand{\cH}{{\cal H}}

\newcommand{\cD}{{\cal D}}

\newcommand{\cM}{{\cal M}}
\newcommand{\bi}{\bigskip}
\newcommand{\no}{\noindent}
\newcommand{\hk}{\hspace{0.1cm}}
\newcommand{\hs}{\hspace{0.5cm}}

\newcommand{\rk}{\right)}
\newcommand{\lk}{\left(}

\newcommand{\sli}{\sum\limits}
\newcommand{\pli}{\prod\limits}
\newcommand{\il}{\int\limits}
\begin{document}
\renewcommand{\thefootnote}{\fnsymbol{footnote}}
\def\slash#1{#1 \hskip -0.5em / }

\vskip 1.5 truecm

\noindent
\bi

\setlength{\baselineskip}{20pt}
\no
{\large \bf Emergence of the Haar measure in the standard functional
integral representation of the Yang-Mills partition function}
\bi

\noindent
\begin{center}
H. Reinhardt\footnote{Supported by DFG Re 856/1-3}\\
{\it Institut f"ur Theoretische Physik, Universit"at T"ubingen}\\
{\it Auf der Morgenstelle 14, D-72076 T"ubingen, Germany\footnote{permanent
address}} \\
and\\
{\it Center for Theoretical Physics}\\
{\it Laboratory for Nuclear Science}\\
{\it and Department of Physics}\\
{\it Massachusetts Institute of Technology}\\
{\it Cambridge, Massachusetts 02139, USA}
\end{center}  
\bi

\no
\begin{abstract}
\no
The conventional path integral expression for the Yang-Mills transition
amplitude with flat measure and gauge-fixing built in via the
Faddeev-Popov method has been claimed to fall short of guaranteeing
gauge invariance in the non-perturbative regime. We show, however, that
it yields the gauge invariant partition function where the
projection onto gauge invariant wave functions is explicitly performed
by integrating over the compact gauge group. In a variant of maximal
Abelian gauge the
Haar measure arises in the conventional Yang-Mills path integral from the
Faddeev-Popov determinant.
\end{abstract}
\bi

\no
{\large \bf 1. Introduction}
\bi

\no
The importance of gauge invariance for confinement is generally accepted.
Several approaches have been proposed, which explicitly
resolve Gau"s' law to obtain a gauge invariant description either
directly in terms of gauge invariant variables \cite{R1} or at least 
in unconstrained
variables \cite{R2}. It has been argued that the conventional path-integral
expression for quantizing $SU (N)$ Yang-Mills theory \cite{RIZ} falls short of 
guaranteeing
gauge invariance in the non-perturbative regime \cite{R3}.
One has therefore constructed alternative path integral representations
of the Yang Mills transition amplitude where gauge invariance is
guaranteed by explicitly projecting the external states on
gauge invariant states \cite{R4,R5}. Projection onto gauge invariant 
states basically
means integration over the compact gauge group with the corresponding
Haar measure. It is the apparent absence of the Haar measure in the
conventional functional integral representation which has been the main
subject of criticism \cite{R3,R4}.
\bi

\no
In this note we show for the partition function that the conventional
functional integral representation with the gauge fixed by the
Faddeev-Popov method fully respects the gauge invariance and is
therefore also applicable in the non-perturbative regime. In particular
we show that the invariant (Haar) 
measure of the gauge group
is, in fact, contained in the conventional path integral representation
and explicitly arises in certain gauges in the form of the Faddeev Popov
determinant.

\vspace{0.5cm}
\no
To make the paper self-contained and to fix our notation we have first
to summarize some well known facts and put them in the appropriate
context.
\bi

 \no
{\large \bf 2. The Yang-Mills transition amplitude}
\bi

\noindent
We consider (Euclidian) Yang-Mills theory with gauge group $G = SU
(N)$. In the Weyl gauge $A^{a}_{0} (x) = 0$ the dynamical degrees of
freedom (coordinates) are the vector potential $A^{a}_{i}  (x)$, and the
Hamilton operator is defined by
\be
H = \int d^3 x \lk \frac{g^2}{2} E^{a}_{i} (x)  E^{a}_{i} (x) +
\frac{1}{2 g^2} B^{a}_{i} (x)  B^{a}_{i} (x) \rk \hk .
\ee
Here the electric field
$E^{a}_{k} (x) = \frac{1}{i} \hk \frac{\delta}{\delta A^{a}_{i}  (x)}$
represents the canonical momentum conjugate to $A^{a}_{i}  (x)$ and
$B^{a}_{k} (x) = \epsilon_{kij} \lk \partial_i A^{a}_{j} (x) +
\frac{1}{2} f^{abc} A^{b}_{i} A^{c}_{j} \rk $
is the magnetic field. Furthermore $ f^{abc}$ is the structure constant
of
the gauge group and $g$ denotes the bare coupling constant. We also
define the matrix valued fields $A_\mu (x) = A^{a}_{\mu} (x) T^a$ with
$T^a$ being the anti-hermitian generators of the gauge group satisfying
$\left[T^a, T^b \right] = f^{abc} T^c$.
\bi

\no
Let $ | C \rangle$ denote an eigenstate of ${A}_{i}  (x)$, i.e.
\be
{A}_{i} (x) | C \rangle = C_i (x) | C \rangle
\ee
with some classical field function $C_i (x)$, so that each wave
functional $\Psi_k (C)$ can be expressed as $\Psi_k (C) = \langle
C | k \rangle$. The quantity of interest
is the quantum transition amplitude $\langle C' | e^{- HT} | C \rangle$.
Due to the gauge invariance of $H$ the transition amplitude is invariant
under simultaneaous gauge transformations $\Omega \lk \vec{x} \rk$ of
initial and final field configurations
\be
\label{C1}
\langle C'^{\Omega} | e^{- HT} | C^{\Omega} \rangle = \langle
C' |  e^{- HT} | C \rangle 
\ee
where
\be
 C^{\Omega}_{i} = \Omega C_i \Omega^\dagger + \Omega \partial_i
\Omega^\dagger \hs , \hs \Omega \in G
\ee
denotes the gauge transform of $C_i (x)$. But the transition amplitude
is not invariant under a separate (independent) gauge transformation of
one of the external fields.
\bi

\no
The gauge invariant transition amplitude of Yang-Mills theory is
obtained by projecting the external states onto gauge invariant
states \cite{R4,R5}
\be
\label{C2}
Z [C', C ] = \langle C' | e^{- HT} P | C \rangle \hk , 
\ee
where the projector is defined by
\be
\label{Y2}
P | C \rangle = \sli_n e^{- i n \Theta} \il_G \cD \mu (\Omega_n) |
C^{\Omega_n} \rangle \hk . 
\ee
Here $\Theta$ is the 
vacuum angle \cite{R6,R7}, and the functional integration with respect
to the invariant
(Haar) measure of the gauge group, $\mu (\Omega)$, extends over all
time-independent gauge transformations $\Omega_n (\vec{x})$ with winding
number $n$. For a gauge transformation $\Omega (x)$ the winding number
is defined by
\be
\label{Y4}
n [\Omega] = \frac{1}{24 \pi^2} \int d^3 x \epsilon_{ijk} tr L_i L_j L_k
\hs , \hs L_k = \Omega \partial_k \Omega^\dagger \hk . 
\ee
As usual we assume here that the gauge function $\Omega \lk \vec{x} \rk$
approaches a unique value $\Omega_\infty$ for $| \vec{x} | \to \infty$ so
that $R^3$ can be compactified to $S^3$ and $n [\Omega]$ is a
topological invariant. For simplicity we will choose $\Omega_\infty =
1$.
\bi

\no
At each $\vec{x}$ the gauge functions $\Omega (\vec{x}) \in G$ can be
diagonalized, which yields the Cartan decomposition
\be
\label{Y5}
\Omega_n (\vec{x}) = V^{\dagger}_{n}  (\vec{x}) \omega_n  (\vec{x}) V_n
(\vec{x}) \hk , 
\ee
where $\omega_n (x)$ is a diagonal unitary matrix living in the Cartan
subgroup (invariant torus) $H = U (1)^{N - 1}$ and 
$V_n  (\vec{x})$
lives in the corresponding coset $G/H$. Note that this representation is
unique only up to space-dependent Abelian gauge transformations $ v
(\vec{x}) , V_n (\vec{x}) 
\to v (\vec{x})  V_n (\vec{x})$.
More precisely $v (x)$ is an element of the so-called normalizer $\cN$
of $H$ in $G$, and $\cN / H = W$ is the Weyl group, which for $G = SU
(N)$ is given by the group of permutations of $N$ elements, $S_{(N)}$
\cite{R7B}. The integration over the gauge group can then be expressed
by the Weyl formula \cite{R7B}
\be
\label{Z1}
\il_G d \mu \lk \Omega_n \rk f  \lk \Omega_n \rk = \frac{1}{|W|} \il_H
d  \bar{\mu} (\omega_n) \il_{G/H} d V f \lk V^{\dagger}_{n} \omega_n
V_n \rk \hk , 
\ee
where $|W|$ is the order of the Weyl group $\lk |W| = N! \hk {\rm for} \hk
G =
SU (N) \rk$ and
the reduced Haar measure $\bar{\mu} (\omega) $ is defined by \cite{R7A}
\be
\label{Z2}
d \bar{\mu} (\omega) & = & \pli_k d \lambda_k \sli_p \delta \lk \sli_i
\lambda_i - 2 \pi p \rk \pli_{i < j} \sin^2 \frac{\lambda_i -
\lambda_j}{2} \hk . 
\ee
Here $i \lambda_k$ denotes the eigenvalue of $\ln \omega$ (the
$\lambda_k$ are real).
\bi

\no
Let us also mention that the diagonalization (\ref{Y5}), which is possible
pointwise, cannot be done smoothly and globally due to topological
obstructions, i.e. in general for an arbitrary $\Omega_n$ there is no
globally defined smooth $V_n (x)$. As a consequence, when the Weyl formula
(\ref{Z2}) is applied to functional integrals the summation over the
different topological sectors has to be included. 
We will return to this point later.
\bi

\no
Of particular interest is the partition function
\be
\label{Y3}
Z = \int \cD C_i \langle C | e^{- H T} P | C \rangle \hk , 
\ee
which upon using the completeness of eigenstates $|k \rangle$ of $H$,
$ H | k \rangle = E_k | k \rangle $ and $P^2 = P$ can be written
as 
\be
\label{Y1}
Z = \int \cD C_i \sli_k \tilde{\Psi}_k (C)  e^{- E_k T} 
 \tilde{\Psi}^{*}_{k} (C) \hk . 
\ee
Here
\be
\label{X1}
\tilde{\Psi}_k (C) = \langle C | P | k \rangle 
\ee
are the gauge ``invariant'' energy eigenfunctionals (i.e. the gauge
invariant eigenstates of $H$ in the ``coordinate'' representation),
which have been assumed to be properly normalized
\be
\int \cD C_i \tilde{\Psi}^{*}_{k} (C) \tilde{\Psi}_l (C) = \delta_{kl} \hk .
\ee
It is then seen that eq. (\ref{Y1}) reduces, in fact, to the standard
form of the partition function
\be
Z = \sli_k e^{- E_k T} \hk .
\ee
Let us also mention that the wave functionals (\ref{X1}) are invariant
only under ``small'' gauge transformation (with zero winding number). 
For a ``large'' gauge
transformation $\Omega_n$ with winding number $n$ they transform as
\be
\tilde{\Psi}_k \lk C^{\Omega_n} \rk = e^{- i n \Theta} \tilde{\Psi}_k
(C)
\ee
as can be inferred from (\ref{X1}) using the explicit form of $P$
(\ref{Y2}).
\bi

\no
Using the Cartan decomposition  (\ref{Z1}) for the group integration in
$P$ (\ref{Y2}) the partition function (\ref{Y3}) becomes
\be
Z = \sli_n e^{i n \Theta} \il_H \cD \bar{\mu} (\omega_n) \il_{G/H} \cD
V_n \int \cD C_i \langle C | e^{ - HT} | C^{V^{\dagger}_{n} \omega_n
V_n} \rangle \hk .
\ee
Using further the invariance of $H$ under gauge transformations (see eq.
(\ref{C1}))
\be
\langle  C | e^{- HT} | C^{V^{\dagger}_{n} \omega_n
V_n} \rangle = \langle C^{V_n} | e^{- HT} | \lk C^{V_n} \rk^{\omega_n}
\rangle
\ee
and changing the integration variable $C_i \to C^{V}_{i} \lk \cD C_i =
\cD C^{V}_{i} \rk$ the integration over the coset $\int \cD V_n$ becomes
trivial yielding an irrelevant constant, and we obtain
\be
\label{Z3}
Z = {\rm const} \sli_n e^{- i n \Theta} \il_H \cD \bar{\mu} \lk \omega_n
\rk \int \cD C_i \langle C | e^{- HT} | C^{\omega_n} \rangle \hk . 
\ee
\bi

\no
{\large \bf 3. Path integral representation of the Yang-Mills
amplitude}
\bi

\no
Following the standard procedure \cite{R8} one derives the
following functional integral representation of the transition
amplitude
\be
\label{C5a}
\langle C' | e^{- HT} | C \rangle = \il^{C'}_{C} \cD A_i (x) e^{- S_{YM}
\left[ A_0 = 0, A_i \right] } \hk , 
\ee
where the functional integration is performed over all classical field
configurations $A_i (x)$ satisfying the boundary
conditions
\be
\label{C6a}
A_i \lk x_0 = 0, \vec{x} \rk & = & C_i \lk  \vec{x} \rk\nonumber\\
A_i \lk x_0 = T,  \vec{x} \rk & = & C'_i \lk  \vec{x} \rk  
\ee
and
\be
\label{C5}
S_{YM} \left[ A_0, A_i \right] = \frac{1}{4 g^2} \int d^4 x F^{a}_{\mu
\nu} (x)  F^{a}_{\mu
\nu} (x)
\ee
is the standard Yang-Mills action with
\be
 F^{a}_{\mu
\nu} = \partial_\mu A^{a}_{\nu} - \partial_\nu A^{a}_{\mu} + f^{abc}
A^{b}_{\mu} A^{c}_{\nu}
\ee
being the field strength. 
\bi

\no
Inserting (\ref{C5a}) into (\ref{C2}) we obtain
\be
\label{C20}
Z [C', C] = \sli_n e^{- i n \Theta} \il_G \cD \mu (\Omega_n)
\il^{C'}_{C^{\Omega_n}} \cD A_i (x) e^{- S_{YM} \left[ A_0 = 0, A_i
\right]} \hk . 
\ee
For the following it is convenient to introduce the Pontryagin index
(topological charge)
\be
\nu [A] = - \frac{1}{16 \pi^2} \int d^4 x tr \lk F_{\mu \nu}
\tilde{F}_{\mu \nu} \rk \hk ,
\ee
where $\tilde{F}_{\mu \nu} = \frac{1}{2} \epsilon_{\mu \nu \kappa
\lambda}  F_{\mu \nu}$ is the dual field strength. In the $A_0 = 0$
gauge this quantity is related to the Chern-Simons action
\be
S_{CS} [A] = - \frac{1}{8 \pi^2} \int d^3 x \epsilon_{ijk} tr \left\{
A_i \partial_j A_k + \frac{2}{3} A_i A_j A_k \right\}
\ee
of the temporal boundary values of the spatial gauge fields $A_i \lk x_0
= 0 \rk = C^{\Omega_n}_{i} \equiv C''_i, A_i \lk x_0 = T \rk = C'_i$ by
\cite{R5}
\be
\nu [A_0 = 0, A_i] = S_{CS} \left[ C'_i \right] - S_{CS}
\left[ C''_i \right] \hk .
\ee
Furthermore under gauge transformation $\Omega$ the Chern Simons action
transforms as  \cite{R7}
\be
S_{CS} \left[ A^{\Omega}_{i} \right] = S_{CS} \left[ A_i \right] + n
[\Omega] \hk .
\ee
Therefore the winding number $n = n [\Omega_n]$ (\ref{Y4}) in
(\ref{C20}) can be expressed as
\be
n = n \left[ \Omega_n \right] = S_{CS} [C'] - S_{CS} [C] - \nu \left[ A_0 =
0, A_i \right] \hk .
\ee
This representation of $n [\Omega]$ is convenient since the first two
terms on the r.h.s. depend only on the externally given boundary fields
while the last term is gauge invariant.
\bi

\no
Let us now remove the gauge function $\Omega_n$ from the boundary field
$A_i \lk x_0 = 0 \rk = C^{\Omega_n}_{i}$ in eq. (\ref{C20}) 
by performing the following
time-dependent gauge transformation
\be
\label{C21}
U = \Omega^{\frac{x_0}{T} - 1}_{n} \hk . 
\ee
Using the gauge invariance of $S_{YM} [A]$ and $\nu [A]$ we find after a
change of integration variables $\lk A^{U}_{i} \to A_i \rk$
\be
\label{C22}
Z [C', C] & = & \exp \left[ - i \Theta \lk S_{CS} [C'] - S_{CS}
[C] \rk \right]\nonumber\\
& & \sli_n \int \cD \mu \lk \Omega_n \rk \il^{C'_i}_{C_i} \cD A_i e^{-
S_{YM} \left[ A_0, A_i \right] + i \Theta \nu \left[ A_0, A_i \right]}
\hk , 
\ee
where a time-independent temporal gauge potential
\be
\label{C22a}
A_0 = - \frac{1}{T} ln \Omega_n 
\ee
has been induced by the gauge transformation (\ref{C21}) and the
functional integration is now performed with the boundary conditions $A_i
\lk x_0 = 0, \vec{x} \rk = C_i \lk \vec{x} \rk, A_i \lk x_0 = T, \vec{x}
\rk = C'_i \lk \vec{x} \rk$. Note that the winding number of the
original gauge function $\Omega_n \lk \vec{x} \rk$ is now encoded in the
temporal gauge potential (\ref{C22a}). By integrating over all possible
(time-independent) field configurations $A_0 \lk \vec{x} \rk$ 
compatible with the boundary
condition $\Omega_n \lk | \vec{x} | \to \infty \rk = 1$,
i.e.\footnote{Without loss of generality we can restrict ourselves to
a single Riemann sheet of the logarithm, e.g. $p = 0$.}
$A_0 \lk | \vec{x} | \to \infty \rk = 2 \pi p \hk i \hk , \hk p - {\rm
integer,}$
we will automatically include the summation over all winding numbers
$n$. We will henceforth omit this sum and understand that it is included
in the (functional) integral over $A_0 \lk \vec{x} \rk$\footnote{Let us
also mention that eq. (\ref{C22}) is equivalent to the functional
integral representation of the gauge invariant transition amplitude
derived in ref. \cite{R4} in a somewhat different way. We would have
obtained that representation had we parametrized $\Omega_n = \Omega U_n$
with $n [\Omega] = 0$ and chosen $\Omega$ instead of $\Omega_n$ in eq.
(\ref{C21}).}.
\bi

\no
For the partition function (\ref{Z3}) we obtain from (\ref{C22}) the
representation
\be
\label{C23}
Z_\Theta = \int \cD \bar{\mu} \lk \omega \rk 
\il_{\stackrel{\rm periodic}
{\rm b.c.}} \cD A_i (x) e^{- S_{YM} \left[ A_0, A_i \right] + i \Theta \nu
\left[ A_0, A_i \right]} \hk , 
\ee
where
the induced temporal gauge potential
\be
\label{C26}
A_0 = - \frac{1}{T} ln \omega \in \cH 
\ee
is now diagonal, i.e. lives in  the Cartan subalgebra and the functional
integration runs over periodic spatial fields $A_i \lk x_0 = T, \vec{x}
\rk = A_i \lk x_0 = 0, \vec{x} \rk$.
\bi

\no
Eq. (\ref{C23}) should be compared with the standard representation of
the Yang-Mills partition function, which is used for 
finite temperature QCD considerations \cite{R8A}
\be
\label{C24}
Z_\Theta =  \il_{\stackrel{\rm periodic} 
{\rm b.c.}} \cD A_\mu  \delta_{gf.} (A) e^{- S_{YM} [A] + i \Theta \nu
[A]} \hk . 
\ee
Here the integration is over all temporally periodic field
configurations 
\be
\label{C25}
A_\mu \lk x_0 = T \rk = A_\mu \lk x_0 = 0 \rk  
\ee
and
$\delta_{gf.} (A) = \delta \lk \chi (A) \rk \det \cM$ denotes the
gauge-fixing according to the Faddeev-Popov method with $\chi (A) = 0$
being the gauge condition and $\det \cM$ the corresponding
Faddeev-Popov determinant. 
\bi

\no
Eq. (\ref{C23}) is ``almost'' the
standard representation (\ref{C24} )
except for the presence of the invariant Haar measure for $A_0$ in 
(\ref{C23}).
Furthermore $A_0$ is diagonal and time-independent in  (\ref{C23}) but 
time-dependent
in (\ref{C24})\footnote{An alternative
functional integral representation for the gauge invariant transition
amplitude with a time-dependent $A_0$ was also
derived in ref. \cite{R4}. This
representation is, however, equivalent to eq. (\ref{C22}). Let us also
mention that representation (\ref{C22}) follows also from the lattice
formulation by using the gauge $\partial_0 U_0 \lk \vec{x}, x^0 \rk = 0$
(where $U_0 (x)$ denotes the temporal link) 
and taking the continuum limit in the spatial directions.} and there is
no explicit gauge fixing included in (\ref{C23}).
\bi

\no
In ref. \cite{R3,R4} the importance of the Haar measure was emphasized, which
ensures gauge invariance and (in the presence of quarks) central
symmetry. (The latter leads naturally to quark confinement \cite{R4}.) 
It is usually argued that the use of the flat integration measure is
justified only for perturbation theory since the Haar measure 
reduces to the flat
measure near $A_0 = 0$.
We will now show, however, that for the partition function both 
representations (\ref{C23}) and (\ref{C24} )
are completely equivalent,
even in the non-perturbative regime. 
\bi 

\no
{\large \bf 4. Equivalence proof}
\bi

\no
To reduce (\ref{C24}) to (\ref{C23}) (which uses a time-independent $A_0$
) we choose the gauge
\be
\label{G6}
\partial_0 A_0 = 0 \hk , 
\ee
which is compatible with the periodic boundary condition (\ref{C25}). This
constraint still allows for time-independent gauge transformations $V
\lk \vec{x} \rk$,
 under which $A_0$ in (\ref{C24})
transforms homogeneously
\be
\label{G7}
A_0 (\vec{x}) \longrightarrow A^{V}_{0} (\vec{x}) = V (\vec{x}) 
A_0 (\vec{x}) V^\dagger (\vec{x})  \hk . 
\ee
Therefore this residual gauge freedom can be exploited to diagonalize
$A_0 (x) $, which 
implies the gauge condition
\be
\label{G8}
A^{ch}_{0} (\vec{x}) = 0 \hk , 
\ee
where $A^{ch}_{0}$ is the off-diagonal (charged) part of $A_0$. Again
the diagonalization can be done pointwise but in general 
not globally and smoothly
due to topological obstructions. We face here the same problem for the
Lie algebra as we observed in the Cartan decomposition of the gauge
group (see the discussion following eq. (\ref{Y5})). At certain points
$\vec{x} = \vec{x}_S$ in space the field $A_0  (\vec{x})$ is
degenerate (i.e. two eigenvalues coincide) such that the unitary matrix $V
(x) \in G$ in (\ref{G7}) necessary to make $A^V \lk \vec{x} \rk$
diagonal is not well defined, i.e. at these points the
constraint (\ref{G8}) does not fix the gauge uniquely.
(For $G = SU (2)$ the degeneracy points are those, where the gauge
function $A_0 (x)$ vanishes.) The physical
meaning of these degeneracy points is easily understood by observing
that eqs. (\ref{G6}), (\ref{G8}) are a variant of maximal 
Abelian gauge \cite{R9}.
Therefore by following 't Hooft's arguments \cite{R9}
one easily shows that the degeneracy points $\vec{x} = \vec{x}_S$ 
are the positions of
magnetic monopoles in the spatial gauge potential $A_i (x)$.
In fact, under a space-dependent gauge transformation $V (\vec{x}), A_i (x)$
transforms inhomogeneously
\be
A^{V}_{i} = V A_i V^\dagger + V \partial_i V^\dagger \hk .
\ee
In the vicinity of a degeneracy point $\vec{x} = \vec{x}_S$ of $A_0 \lk 
\vec{x} \rk$, the structure of $V (x)$ is such 
that the Abelian part of the
inhomogeneous term $V \partial_i V^\dagger$ develops a
magnetic monopole at $\vec{x} = \vec{x}_S$.
\bi

\no
The gauge conditions (\ref{G6}) and (\ref{G8})
still remain invariant under time-independent Abelian gauge
transformations. To prove the equivalence between (\ref{C23}) and
(\ref{C24}) we need not fix this residual gauge since the same
residual gauge freedom is also present in eq. (\ref{C23})\footnote{
This residual gauge freedom could be fixed e.g. by the Coulomb type of gauge
condition
\be
\label{G9} 
\il^{T}_{0} d x^0 \vec{\nabla} \vec{A}^n = 0 \hk ,  
\ee
where $A^{n}_{i}$ is the diagonal (neutral) part of $A_i$. Inclusion of
this gauge condition would, however, not change the Faddeev-Popov
determinant (see eq. (\ref{G49}) below). }
Note also, that for the charged part $A^{ch}_{0}$ the condition (\ref{G6}) is
already included in (\ref{G8}), so we have to implement (\ref{G6}) only
for
the neutral (diagonal) part $A^{n}_{0}$ of $A_0$, which lives
in the Cartan subgroup $(U (1))^{N - 1}$.
Therefore our gauge conditions read
\be
\label{G10}
\chi^{a_0} (x) & = & \partial_0 A^{a_0}_{0} (x) = 0 
\\
\label{G11}
\chi^{\bar{a}} (x) & = & A^{\bar{a}}_{0} (x) = 0 
\ee
Here and in the following we use the indices $a = a_0$ and $a = \bar{a}$ 
for diagonal and off-diagonal
generators, respectively.
\bi

\no 
The gauge fixing constraints (\ref{G10}) and (\ref{G11}) give rise to a
Faddeev-Popov kernel
\be
\cM^{a b_0} (x, y) & = & \hat{D}^{a b_0}_{0} (x) \partial^{y}_{0}
\delta^{(4)} (x, y) 
\hk ,
\nonumber\\
\cM^{a \bar{b}} (x, y) & = &  \hat{D}^{a \bar{b}}_{0} (x) \delta^{(4)}
(x, y) \hk ,
\ee
where
\be
\hat{D}_{\mu} = \partial_\mu + \hat{A}_\mu \hs , \hs  \hat{A}_\mu =
A^{a}_{\mu} \hat{T}^a
\ee
is the covariant derivative with $ \hat{T}^a$ being the generators in
the adjoint representation $\left[ \lk  \hat{T}^a \rk^{bc} = - f^{abc}
\right]$. Using $f^{a b_0 c_0} = 0$, for gauge configurations satisfying
the gauge constraints (\ref{G10}),  (\ref{G11}) the Faddeev-Popov kernel
reduces to
\be
\label{G13}
\cM^{ab}  \equiv  \lk
\begin{array}{cc}
\cM^{a_0 b_0} & \cM^{a_0 \bar{b}} \\
\cM^{\bar{a} b_0} & \cM^{\bar{a} \bar{b}}
\end{array}
\rk 
 =  \lk \begin{array}{cc}
- \delta^{a_0 b_0} \partial^{x}_{0} \partial^{x}_{0} \delta^{(4)}
(x, y) 
& 0 \\
0 & \hat{D}^{\bar{a} \bar{b}}_{0} (x)  \delta^{(4)}
(x, y)
\end{array} \rk \hk .   
\ee
Since this matrix is block-diagonal we find for the Faddeev-Popov
determinant
\be
\label{G49}
Det \cM = {\rm const} Det \lk \hat{D}^{\bar{a} \bar{b}}_{0} (x)
\delta^{(4)} (x, y) \rk \hk , 
\ee
where the irrelevant constant arises from the Cartan subgroup (upper
left block in (\ref{G13})). 
\bi

\no
Due to the adopted gauge, eqs. (\ref{G6}) and  (\ref{G8}), the eigenvalue
equation
\be
i \hat{D}^{\bar{a} \bar{b}}_{0} \phi^{ \bar{b}} \equiv \lk i \partial_0
\delta^{\bar{a} \bar{b}} + \hat{A}^{\bar{a} \bar{b}}_{o} (\vec{x}) \rk
\phi^{\bar{b}} = \mu \lk \vec{x} \rk \phi^{\bar{a}}
\ee
is easily solved. In the fundamental representation $\lk \phi =
\phi^{\bar{a}} T^{\bar{a}} , T^a = - i \frac{\lambda^a}{2} \rk$ this
equation reads
\be
i \partial_0 \phi + \left[ i A_0  (\vec{x}), \phi \right] = \mu \phi
\hs , \hs A_0 = A^{c_0}_{0} T^{c_0} \hk .
\ee
Adopting the Weyl basis, in which the index $\bar{c}$ for the
off-diagonal generator $T^{\bar{c}}$ is expressed by the two respective
indices $(k, l)$ of the fundamental representation for which
$T^{\bar{c}}_{kl} = - T^{\bar{c}^\dagger}_{lk} \neq 0$, i.e. $\bar{c} =
(k,l)$, the eigenvalues are given for the temporally periodic 
boundary condition 
 (\ref{C25}) by
\be
\mu_{n, \bar{c}} (\vec{x}) = \omega_n + i \left[ \lk A_0  (\vec{x}) 
\rk_{kk} -  \lk A_0  (\vec{x}) \rk_{ll}
\right] \hs , \hs \omega_n = \frac{2 \pi n}{T} \hk .
\ee
%
Using
\be
\sin x = x \pli^{\infty}_{n = 1} \lk 1 - \lk \frac{x}{\pi n} \rk^2 \rk
\ee
straightforward evaluation yields
\be
\label{G50}
Det \cM & = & {\rm const.} \pli^{\infty}_{n = - \infty} \pli_{k \neq l}
\mu_{n, (k,l)}\nonumber\\
& = &  {\rm const.} \pli_{k > l} \sin^2 \frac{T \lk \lk i A_0 \rk_{kk} -
\lk i A_0 \rk_{ll} \rk}{2} \hk . 
\ee
Taking into account that by definition of $A_0$ (\ref{C26}) the integration
over $A_0$ extends from $- \infty$ to $\infty$ and furthermore $tr A_0 =
\sli_k (A_0)_{kk} = i 2 \pi n, n = 0, \pm 1, \cdots$ the Faddeev-Popov 
determinant (\ref{G50}) gives
precisely the (reduced) Haar measure (\ref{Z2}). 
\bi

\no
Therefore in the gauges (\ref{G6}, \ref{G8}) we have
\be
\int \cD A_0 \delta (\chi) Det \cM \cdots = \int \cD \bar{\mu}
 (\omega) \cdots \hs , \hs \omega
 = 
e^{- T A_0} \in H \hk ,
\ee
which shows, that in this gauge 
the functional integral representation (\ref{C24}) coincides with the
representation (\ref{C23}).
Furthermore, the usual functional integral representation  (\ref{C24}) is
invariant under a change of the gauge condition. Therefore, if eq.
(\ref{C24}) reproduces the invariant partition function (\ref{C23}) in
one gauge it does so in any gauge.

\vspace{0.5cm}
\no
One may argue here that the Haar measure escapes in the gauge $A_0 = 0$.
However,
the gauge condition $A_0 = 0$ conflicts with the periodic boundary
condition (\ref{C25}). This can be easily seen by considering the Polyakov line
operator
\be
\label{GX}
L_0 (x) = P \exp \lk \oint\limits_x d x'_0 A_0 \lk x'_0 \vec{x} \rk \rk
\hk , 
\ee
where $P$ denotes path ordering and the integration runs from a point $x
= \lk x_0, \vec{x} \rk$ along the $0$-axis to the point $x = \lk x_0 +
T, \vec{x} \rk$. Due to the periodic boundary condition on $A_0$ the
integration in (\ref{GX}) runs over a closed loop but nevertheless due
to the path-ordering $L_0 (x)$ depends on the starting point $x$. Under
gauge transformation this quantity transforms as 
\be
L_0 (x) \longrightarrow L^{\Omega}_{0} (x) = \Omega (x) L_0 (x)
\Omega^\dagger (x)
\ee
and one can obviously choose a gauge in which $L_0 (x)$ is diagonal
\be
L^{\Omega}_{0} (x) = e^{a_0 (x) T} \hs , \hs a_0 (x) = a^{c_0}_{0}
T^{c_0} \hk .
\ee
But it is impossible to gauge transform $L_0 (x) $ to $L_0 (x) \equiv
1$\footnote{ This can be also easily seen in the lattice formulation. Starting at
$x_0 = 0$ one can bring the links $U_0 (x) = \exp \lk - a A_0 (x) \rk$
to the gauge $U_0 (x) = 1$ except for the last link terminating at $x_0
= T$, which cannot be gauged away due to the periodic boundary
condition.}.
\bi

\no
To summarize we have shown that the usual functional integral
representation with flat integration measure and the gauge fixed 
by the Faddeev Popov method yields
the proper gauge invariant partition function and is hence not
restricted to the perturbative regime contrary to what is commonly
believed. In certain gauges the compact integration measure of the
gauge group arises directly from the Faddeev-Popov determinant.
\bi

\no
Finally let me comment on the degeneracy points of the field $A_0
(\vec{x})$, where the maximal Abelian gauge (\ref{G8}) is not well
defined and monopoles arise in $A_i (x)$. At these
points two of the eigenvalues $\lk A_0 {\lk \vec{x} \rk} \rk_{kk}$
coincide
and, consequently, the Faddeev-Popov determinant (\ref{G50})
vanishes, as usual when the gauge is not unique. 
The field configurations of vanishing Faddeev-Popov determinant define
the Gribov horizon. We may thus conclude that in this context 
the Gribov horizon is built
up from monopole configurations.
\bi

\no
{\large \bf Acknowledgements:}
\bi

\no
This work was largely performed during a visit at MIT. The author
thanks the collegues of the Center for Theoretical Physics, in particular
K. Johnson, for many interesting discussions and also for the
hospitality extended to him. He also acknowledges discussions with M.
Quandt.

\bi

\newpage
\section*{\normalsize \bf References}

\begin{thebibliography}{99}
\bibitem{R1}
K. Johnson, ``The Yang-Mills Ground State'', in `QCD - 20 Years Later',
Aachen, June 1992\\
D.Z. Freedman, P.E. Haagensen, K. Johnson and J.I. Latorre, Nucl.
Phys.\\
P.E. Haagensen and K. Johnson, Nucl. Phys. {\bf B439 } (1995) 597\\
M. Bauer, D.Z. Freedman and P.E. Haagensen, Nucl. Phys. {\bf B428}
(1994) 147
\bibitem{R2}
J. Goldstone and R. Jackiw, Phys. Lett.
{\bf B74} (1978) 81 \\
V. Baluni and B. Grossman, Phys. Lett. {\bf B78} (1978) 226\\
A.G. Izergin et al., Theor. Math. Phys. {\bf 38} (1979) 3\\
Yu. A. Simonov, Sov. J. Nucl. Phys. {\bf 41} (1985) 835, 1014\\
F. Lenz, H.W.L. Naus and M. Thies, Ann. Phys. {\bf 233} (1994) 314\\
N.H. Christ and T.D. Lee, Phys. Rev. {\bf D22} (1980) 939\\
J. Gervais and B. Sakita, Phys. Rev. {\bf D18} (1978) 453\\
M. Creutz, I.J. Muzinich and T.N. Tudron, Phys. Rev. {\bf D19} (1979)
531
\bibitem{RIZ}
C. Itzykson and J.-B. Zuber, Quantum Field Theory, McGraw-Hill, 1980
\bibitem{R3}
G.C. Rossi and M. Testa, Nucl. Phys. {\bf B163}
(1980) 109, {\bf B179} (1980) 477\\
J. Polonyi, Phys. Lett. {\bf B213} (1988) 340
\bibitem{R4}
K. Johnson, L. Lellouch and J. Polonyi, Nucl. Phys. {\bf B367} (1991) 675
\bibitem{R5}
M. L"uscher, R. Narayanan, P. Weisz and U. Wolff,
                Nucl. Phys. {\bf B384} (1992) 168 
\bibitem{R6}
R. Jackiw and C. Rebbi, Phys. Rev. Lett. {\bf 37} (1976) 172\\
C.G. Callan, R.F. Dashen and D.J. Gross, Phys. Lett. {\bf B63} (1976)
334
\bibitem{R7}
R. Jackiw, Rev. Mod. Phys. {\bf 52} (1978) 661
\bibitem{R7A}
H. Weyl, The classical groups, Princeton University Press, Princeton, NJ,
1973
\bibitem{R7B}
T. Br"ocker and T. tom Dieck, Representations of Compact Lie Groups,
Springer Graduate Texts in Mathematics, New York, 1985
\bibitem{R8}
R. Feynman and A. Hibbs, Quantum Mechanics and Path integrals,
McGraw-Hill, New York, 1965
\bibitem{R8A}
D.J. Gross, R.D. Pisarski and L.G. Yaffe, Rev. Mod. Phys. {\bf 53}
(1981) 43
\bibitem{R9}
G 't Hooft, Nucl. Phys. {\bf B190} (1981) 455

\end{thebibliography}
\end{document}